\documentclass[aps,twocolumn]{revtex4}
\usepackage{graphicx}
\usepackage{amssymb}

\begin{document}
\title{Suppression of 2D superconductivity by the magnetic field:
 quantum corrections
 vs superconductor-insulator transition.}
\author{V.F.Gantmakher}
\author{S.N.Ermolov}
\author{G.E.Tsydynzhapov}\email{gombo@issp.ac.ru}
\author{A.A.Zhukov}
\affiliation{Institute of Solid State Physics RAS, Chernogolovka, Russia}
\author{T.I.Baturina}
\affiliation{Institute for Semiconductor Physics SB RAS, Novosibirsk, Russia}
\begin{abstract}
Magnetotransport of superconducting Nd$_{2-x}$Ce$_x$CuO$_{4+y}$ (NdCeCuO) films
is studied in the temperature interval 0.3--30\,K. The microscopic theory of
the quantum corrections to conductivity, both in the Cooper and in the
diffusion channels, qualitatively describes the main features of the experiment
including the negative magnetoresistance in the high field limit. Comparison
with the model of the field-induced superconductor--insulator transition is
included and a crossover between these two theoretical approaches is discussed.
\end{abstract}
 \maketitle

The superconductor--insulator transition (SIT) is an example of the
quantum phase transitions\,\cite{Sondhi} which constitutes drastic change
of the ground state of the system at zero temperature with varying a
parameter. The field was pioneered by A.\,Goldman {\it et al.} in the
1989 \cite{Bi} who obtained the transition from  insulating to
superconductive state in the thin Bi film with the change of its
thickness. Later, Fisher \cite{Fisher} suggested  existence of magnetic
field-induced SIT in two-dimensional (2D) systems and Hebard and Paalanen
demonstrated \cite{Hebard,Paal} such a transition in amorphous InO$_x$
films. Numerous results obtained in several other materials by different
groups \cite{Kapitulnik,Okuma,Tanda,NCCO} were also interpreted within
the framework of the field-induced SIT. Main arguments in favor of this
interpretation were negative derivative of resistance $\partial
R/\partial T$ in the fields above the critical and existence of a
finite-size scaling, i.e. existence of some critical region on the
($T,B$)--plane where the behavior of the system was governed by
competition of the quantum phase transition correlation length
$\xi\propto (B-B_c)^{-\nu}$ and thermal length $L_T\propto T^{1/z}$ with
$z$ and $\nu$ being constants called the critical exponents. All relevant
quantities in this region are supposed to be universal functions $f$ of
ratio of  the lengths which can be written in the form of scaling
variable $(B-B_c)/T^{1/z\nu}$. For the resistivity in two dimensions
$R_\square$ this dependence takes form\,\cite{Fisher}
\begin{equation}
\label{scaleq} R_\square(B, T)=R_c f[(B-B_c)/T^{1/z\nu}],
\end{equation}
where $R_c$ is a constant of the order of $h/4e^2\approx6.5$\,k$\Omega$.
It is called the critical resistance.

In the analysis of the experiments \cite{Kapitulnik,Okuma,Tanda,NCCO} the
negative derivative $\partial R/\partial T$ was rated as an indicator of
the insulating state. However, that is not enough: the characteristic of
any insulator is the exponential temperature dependence of the
resistance. This was demonstrated only in InO$_x$ films \cite{ourInO}.
The growth of the resistance with decreasing temperature on the
non-superconducting side of the field-induced transition in the
experiments with MoGe \cite{Kapitulnik}, MoSi \cite{Okuma} and NdCeCuO
\cite{Tanda,NCCO} was minuscule, about ten percent at its best. It
reminded more a metal with quantum corrections to its conductivity than
an insulator. Usually, the authors do not dwell on the issue, considering
weak localization-like behavior to be telltale sign of insulator\,--- as,
according to scaling hypothesis \cite{4band}, there is no
non-superconducting delocalized state at zero temperature in 2D and weak
localization is expected to transform sooner or later into strong.
However, this crossover might be postponed to extremely low temperature
which would never be achieved in practice.

There exists one more sign of SIT. According to the boson--vortex duality
model \cite{Sondhi,Fisher}, the insulating state which appears as the
result of SIT is rather specific; it contains pair correlations between
the localized electrons as the remnant of the superconducting pairing.
Such insulator is called the Bose-insulator \cite{Paal} and the
correlated electrons are called localized electron pairs. These
correlations should be destroyed by strong magnetic field leading to
increase of the carrier mobility, to the negative magnetoresistance
\cite{Shep1} and even to a reentrant insulator--normal-metal transition
\cite{ourInO}. The negative magnetoresistance was observed in MoSi
\cite{OkumaNMR} and NdCeCuO \cite{NCCO}. But it was much weaker than in
InO, just the same as the growth of the resistance with decreasing
temperature discussed above.

When comparing the whole set of data in InO
\cite{Hebard,Paal,ourInO,ourInO2} with those in MoGe \cite{Kapitulnik},
MoSi \cite{Okuma} and NdCeCuO \cite{NCCO}, one can't help impression that
they have many similar features though of different scales of magnitude.
At the same time, it was shown in a set of InO$_x$ films with various
oxygen content $x$ that in low-resistivity films a transition to the
metallic state substitutes SIT, the rate of the temperature dependence
scales down and the whole pattern of curves approaches that of the usual
superconducting transition \cite{ourInO,ourInO2}. The main idea of this
paper follows from this observation. It is to compare experimental set of
data of a ``small-scale'' type with the theory of the superconducting
transition in dirty limit and, keeping in mind its features related to
SIT, to build a bridge between SIT and thermodynamic superconducting
transition.

Experiment was performed on the 1000\AA-thick films of
Nd$_{2-x}$Ce$_x$CuO$_{4+y}$ (NdCeCuO) obtained by laser ablation with
CuO$_2$ planes parallel to the plane of the film. Films were not
superconductive as-grown. In order to obtain superconductivity they were
annealed at 720$^\circ$C in the flowing $^4$He gas for several hours. As
we aimed to study vicinity of the SIT, we were not trying to reach
maximal $T_c$ of this material, but were paying attention for smoothness
and width of the zero-field transition. A sample was chosen with
zero-field transition temperature $T_{c0}=11.8\pm0.4$\,K (found by
fitting of the superconducting fluctuation contribution to the
conductivity above $T_{c0}$) and the transition width $\Delta
T\simeq2$\,K.

The resistivity was measured in the ab-plane by the 4-terminal technique. Both
current and potential probes were attached on the surface of the films by
silver paste. Distance between potential probes corresponded to one square.
Magnetic field was applied perpendicular to the film plane (along c-axis).
Data, both as a function of field at constant temperature and as a function of
temperature at constant field, were obtained though only the latter will be
presented below. The upper panel of Fig.\ref{exp} presents an overview of the
impact of the field on $R(T)$ dependence and the lower one zooms in on the
region of interest, i.e. on the low temperature and high field region.

\begin{figure}
 \includegraphics[width=\columnwidth, clip]{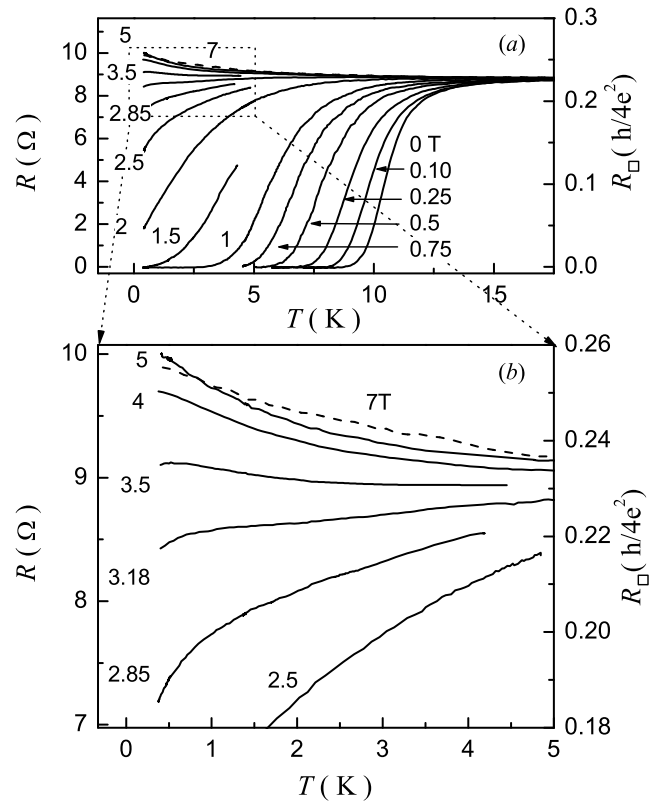}
  \caption{Low-temperature resistivity data for the NdCeCuO film. Panel
  {\it b} is an expansion of the designated area. Curve at 7\,T (dash line)
  is crossing the other ones manifesting the negative magnetoresistance
  below 1\,K.}\label{exp}
\end{figure}

On the right axis of Fig.\,\ref{exp} the resistance reduced per one
CuO$_2$ plane per square is denoted. As NdCeCuO is highly anisotropic
\cite{anisotropy}, it is reasonable to assume the film to be a stack of
2D conducting CuO$_2$ planes with interplane spacing 6\AA,
quasi-independent and connected in parallel. This is supported by
observations of 2D character of quantum interference corrections
\cite{Ponomarev2D} and magnetoresistance \cite{Kussmaul2D}. Later on we
continue discussion in terms of this variable, disregarding full
resistivity and actual thickness of the film. As one could see from
Fig.\,\ref{exp}, the value of the resistance per layer stays quite far
from the quantum resistance $h/4e^2$ expected for the SIT.

The data are quite typical for the material (cf., for example, Ichikawa
{\it et al.} \cite{NCCO}). In the low field region, the transition is
shifted to the lower temperature as the field increases while the shape
of the transition is preserved relatively well.  Above 2\,T, the
transition broadens drastically and eventually disappears; at about
3.5\,T, the $dR/dT$ changes its sign. At higher fields, above 5\,T, the
resistance starts to decrease with the increasing field; it follows from
the crossing of the 5\,T and 7\,T curves that a region of the negative
magnetoresistance exists below 0.8\,K and at $B>5$\,T.

The set of curves $R(T)$ on the lower panel of Fig.\,\ref{exp} is similar
to those obtained in  \cite{Kapitulnik,Okuma,Tanda,NCCO} which had been
regarded as a field-induced SIT. Low-field curves (which bend down) may
be supposed to reach zero resistivity at zero temperature and to become
superconductor, high field curves (which bend up) may be supposed to
diverge toward zero temperature and become insulator. In between, there
is a curve which is almost horizontal; it manifests itself as common
crossing point of all isotherms on the $R-B$ graph. The corresponding
state should be considered as the critical one with the
temperature-independent resistance at the critical field
$B_c\approx3.5$\,T. But instead of seeking scaling parameters we shall
compare experimental data with the microscopic theory of the
superconducting transition in dirty limit formulated in terms of quantum
corrections to the classical Drude conductivity $\sigma_0=e^2/h(k_F l)$
where  $k_F$ is the Fermi wavevector and $l$ is the elastic mean free
path. This comparison became executable due to recent progress in
calculation of the corrections due to superconducting fluctuations
\cite{Galitsky}.

\begin{figure}
\includegraphics[width=\columnwidth,clip]{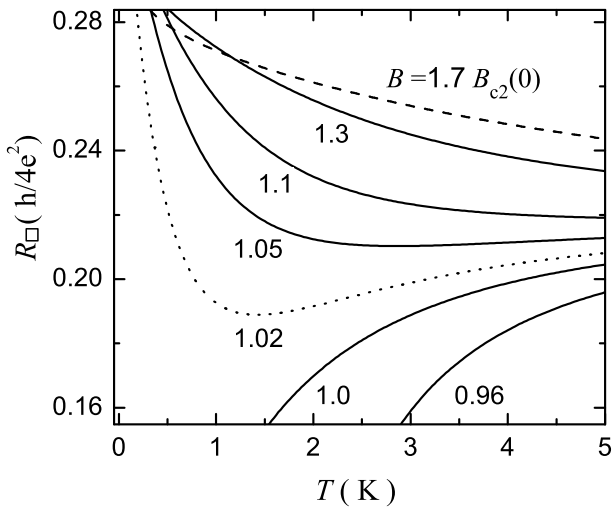}
\caption{Functions $R(T)$ at different $B$ calculated from Eqs.\,(\ref{GL}) and
(\ref{calc}). The curves are labelled by reduced field values. The curve which
shows the negative magnetoresistance is marked out by dash line. The dotted
curve should not be compared to experiment, see text.}\label{GaL}
\end{figure}

All quantum corrections fall into two categories\,--- a one-particle
correction, usually called weak localization, and those due to e-e
interactions. The latter are divided into a diffusion channel correction (also
known as Aronov--Altshuler term) and Cooper channel corrections (also known as
superconductive fluctuations corrections which include Aslamazov--Larkin,
Maki--Thompson and DOS terms). Weak localization and Aronov--Altshuler
corrections diverge at $T\to0$, Cooper channel corrections diverge at $T\to
T_c(B)$ with $T_c(B)$ being mean field transition temperature. When the
superconductivity is suppressed by the magnetic field, $T_c(B)\to0$ and all
corrections are important.

Recently, Galitski and Larkin\,\cite{Galitsky} succeeded in extending
calculations in the Cooper channel for two-dimensional superconductors  to the
low temperature $T\ll T_{c}(0)$ and high magnetic field $B\gtrsim B_{c2}(0)$.
The correction to the conductivity in the dirty limit $\delta\sigma$ is
obtained as the sum of contributions of ten Feynman diagrams in the first
(one-loop) approximation and can be written in the form
\begin{equation}\label{GL}
  \delta\sigma=
  \frac{4e^2}{3\pi h}\left[-\ln\frac rb-\frac{3}{2r}+\psi(r)
  +4(r\psi'(r)-1)\right],
\end{equation}
where $r=(1/2\gamma^\prime)(b/t)$, $\gamma^\prime=e^\gamma=1.781$  is the
exponential of  Euler's constant, and $t=T/T_{c0}\ll1$ and
$b=(B-B_{c2}(T))/B_{c2}(0)\ll1$ are reduced temperature and magnetic field.

To compare these calculations with the experiment, we added to the correction
(\ref{GL}) an additional term to account for Aronov-Altshuler contribution,
which is assumed to be field independent. Weak localization was omitted because
we are interested in the region of rather strong magnetic fields where this
correction was expected to vanish. Finally, we arrived at the formula
\begin{equation}
\label{calc}
 R_{_\square}^{-1}(B,T)=
 \sigma_0+\delta\sigma(B,T)-\alpha\frac{e^2}{h}\ln(T/T^*).
\end{equation}
Inserting $T_{c0}=11.8$\,K and experimental value of the classical conductivity
$\sigma_0=1/R_{_\square}$(7\,T,\ 20\,K), and choosing $T^*=20$\,K to make the
last term zero at 20\,K and $\alpha=1/2$ to match the temperature dependence of
the experimental curve at 7\,T, we get the plot of Fig.\,\ref{GaL} which can be
compared with the experimental one, Fig.\,\ref{exp}b. (Note, that on
Fig.\,\ref{GaL} curves are labelled by reduced field values, those in units of
$B_{c2}(0)$. The same can't be done on Fig.\,\ref{exp}, because experimental
value of $B_{c2}(0)$ is a bit uncertain.)

As one can see, the picture bears clear resemblance to the experiment --- there
is separation between low-field curves which  ``bend down'', and high-field
which ``bend up''; there is also high field negative magnetoresistance at low
temperature. There are two remarkable points: i) the scales of variation of
resistance both with temperature and magnetic field are correct; ii) the region
and the magnitude of the negative magnetoresistance are in reasonable agreement
with the experiment as well.

However, the similarity is qualitative. It is difficult to make it quantitative
and both the experiment and the theory are responsible for this.

Disadvantage of the experiment is hidden in macro-inhomogeneity of the film. It
follows from Fig.\,\ref{GaL} that small 2\%--3\% changes of $B_{c2}(0)$ lead to
drastic shift in the shape of curves $R(T)$, especially near the critical value
of $B$. Inevitable dispersion of the values of $B_{c2}(0)$ along the film
smoothes the curves and clears away the extremum. Hence, one should scarcely
expect to find in the experimental assortment of curves one similar to the
theoretical curve labelled 1.02 (plotted by the dotted line on
Fig.\,\ref{GaL}).

\begin{figure}
\includegraphics[width=\columnwidth,clip]{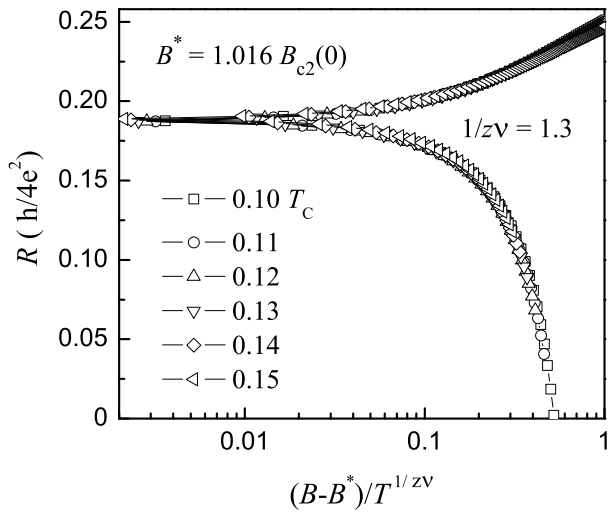}
\caption{``Scaling'' of the curves calculated from Eqs.\,(2) and (3) in
the same way as Fig.\protect\ref{GaL}. Restricted ranges of $T$ and $B$
are selected, see text. }\label{Sca}
\end{figure}

The expression (\ref{GL}) is apparently very sensitive to the function
$B_{c2}(T)$. Basically, this function is an implicit parameter of the theory.
In Ref.\,\onlinecite{Galitsky}, authors used for $B_{c2}(T)$ the mean-field
function from the Werthamer--Helfand--Hohenberg theory. It is doubtful, that
this theory is applicable to high-resistive 2D objects, especially, as the
shape of transition in 2D case should be affected by the vortex motion
(Berezinsky--Kosterlitz--Thouless theory).

As a side note, a comment about the finite-size scaling equation
(\ref{scaleq}) related to SIT. Certainly, expression (\ref{calc}) does not
have form of equation (\ref{scaleq}) and no genuine scaling exists.
However, in a restricted region of values of $T$ and $B$ representation of
the {\it theoretical} curves in the form (\ref{scaleq}) can be done. This
is illustrated by Fig.\,\ref{Sca} where calculated data from the region
$0.98<B/B_{c2}(0)<1.2$ and $0.1<T/T_c<0.15$ are used for the tracing. As
the ``critical'' magnetic field $B^*=1.016B_{c2}(0)$, the crossing point
of several isotherms $R(B)$ was taken; $B^*$ is the field where the
minimum of the isomagnetic curve $R(T)$ is located in the middle of the
chosen temperature region. (Actually, in the limited range of parameters
$B$ and $T$ scaling always exists provided that several curves $R(B)$ have
a common crossing point.) It follows that the scaling tracing is necessary
but not sufficient element of the analysis of the SIT, especially taking
into account that we always deal with the limited temperature range in the
experiment.

Appearance of the negative correction to conductance in the microscopic theory
of the superconductive fluctuations\,\cite{Galitsky} is very remarkable. It
confirms that the superconducting correlations may lead at fields above the
critical one not to the drop but to {\it upsurge} of the resistance. This can
be regarded as the tendency toward the Bose-insulator, which could be
distinguished from the Aronov--Altshuler term because it leads to the negative
magnetoresistance.  All the materials mentioned above can be lined up
demonstrating continuous crossover from the Bose-insulator and gigantic
negative magnetoresistance in InO to faint low-temperature uprise of the
resistance and its tiny drop in strong magnetic fields in MoSi and NdCeCuO. In
essence, these films are similar to each other: they are uniform, highly
disordered films, with the resistance close to quantum value $h/4e^2$.
Nevertheless, experimental observations on InO$_x$ and, for example, on NdCeCuO
are quite different and there is a reason for it.

\begin{figure}
\includegraphics[width=\columnwidth,clip]{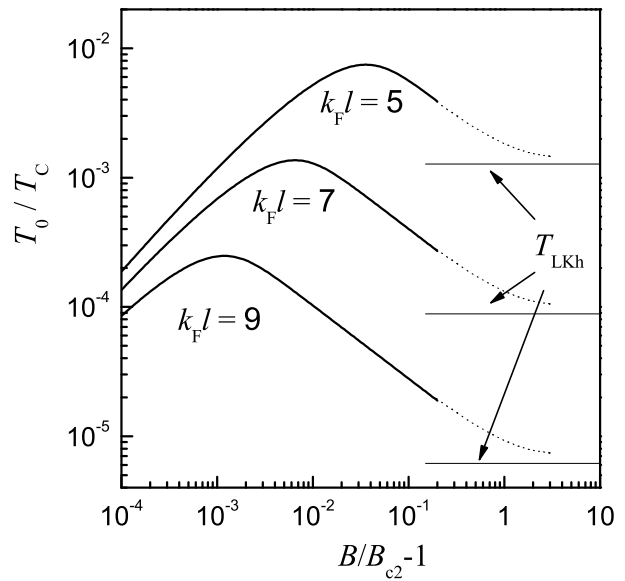}
\caption{Crossover temperature $T_0$ for several reduced values of the
mean free path $l$ calculated by equating to zero the right part of the
Eq.\,(3) for the fields values up to $B=1.2\,B_{c2}(0)$. Dotted lines
qualitatively designate the asymptotic parts of the curves. Levels of
$T_{LKh}$ approximately corresponding to the same values of $l$ are marked
by horizontal lines.}\label{cross}
\end{figure}

There is little doubt that at low enough temperature the growth of $R(T)$ we
observe in the high magnetic field, i.e. in the normal state, will turn
exponential. According to phenomenological estimate suggested by Larkin and
Khmel'nitskii\,\cite{LKh}, the crossover happens when the corrections to the
conductivity reach the level of the conductivity itself. The condition
$\sigma_0\sim(e^2/h)\ln T$ gives crossover temperature
\begin{equation}\label{crossT}
  T_{LKh}\simeq\frac{\varepsilon_F}{k_Fl}e^{-2(k_Fl)},
\end{equation}
where $\varepsilon_F$ and $k_F$ are the Fermi energy and the Fermi wavevector
and $l$ is the elastic mean free path\,\cite{LKh}. Below this temperature there
will definitely be a superconductive state at low field and pronounce
insulating behavior at high field and there would be clear reason to apply SIT
framework. So, the quantum corrections to the conductivity and the quantum
phase transition phenomena are manifested at different temperature regions.

Though $T_{LKh}$ may be very low for normal metal ($T_{LKh}\lesssim1$\,mK
already for $k_Fl\approx5$), there are clear experimental indications
that crossover to bosonic insulator behavior (that is, to the SIT
framework) in the intermediate field range, where pair correlations are
still important, occurs at higher temperature \cite{ourInO}. This is
consistent with theoretical observation \cite{Shep2} that the attractive
interaction stimulates localization by combining single particles into
pairs.

By equating two last terms in the relation (\ref{calc}) to the $\sigma_0$
and solving ensued equation one gets crossover temperature to bosonic
insulator $T_0$ as the function of the magnetic field. These curves for
$\sigma_0$ equal $5\,e^2/h$ (or $k_F l=5$), $7\,e^2/h$ and $9\,e^2/h$ are
presented on Fig.\ref{cross} by solid lines. Thin solid lines present
levels of  $T_{LKh}$ determined by using only last term in the relation
(\ref{calc}) and corresponding $\sigma_0$. As the equation (\ref{GL}) is
valid only in the fields close to $B_{c2}(0)$, the parts of curves in the
higher fields, where $T_0(B)$ approaches $T_{LKh}$, are indicated
qualitatively by dotted lines. In agreement with
Refs.\,\onlinecite{ourInO,Shep2}, the crossover to activation behavior in
the medium-range fields occurs at the temperatures more than order of
magnitude higher than $T_{LKh}$. At the same time, the crossover
temperature falls off exponentially with increasing classical
conductivity so that for the actual value of our experiment it becomes
infinitesimal.  That's why field-induced SIT is so manifest in the
InO$_x$, whereas it is not observed in MoGe or NdCeCuO, and there is no
slightest sign of it in the Al film (note that according to the scaling
hypothesis\,\cite{4band} any metal film should become insulating at $T=0$
if the superconductivity is destroyed by the magnetic field).

To summarize,  we compared experimental data obtained on two-dimensional
NdCeCuO superconductor in magnetic field at low temperature with the
calculations of quantum corrections to the conductivity and found reasonable
agreement. Lack of the activation behavior at high fields  (on the "insulating
side of transition") was the main reason which made inferior comparison of the
same data with the model of field-induced SIT. Apparently, this happened
because the temperature range turned out to be too high for this specific
material. The type of the resistance dependence on the temperature is the guide
in choosing the theoretical approach. To employ framework of the SIT in its
full for NdCeCuO, further essential lowering of the temperature is necessary.

This work was supported by grants RFBR-02-02-16782, RFBR-02-02-08004,
RFBR-03-02-16368 and by grant of Ministry of Science.

\end{document}